# Micropipette aspiration method for characterizing biological materials with surface energy

Y. Ding[1], G. F. Wang[1]*, X. Q. Feng[2] and S. W. Yu[2]

[1]Department of Engineering Mechanics, SVL, Xi'an Jiaotong University, Xi'an 710049, China

[2]Department of Engineering Mechanics, Tsinghua University, Beijing 100084, China

* E-mail: wanggf@mail.xjtu.edu.cn

**Abstract**

Many soft biological tissues possess a considerable surface stress, which plays a significant role in their biophysical functions, but most previous methods for characterizing their mechanical properties have neglected the effects of surface stress. In this work, we investigate the micropipette aspiration to measure the mechanical properties of cells and soft tissues with surface effects. The neo-Hookean constitutive model is adopted to describe the hyperelasticity of the measured biological material, and the surface effect is considered through the finite element method. It is found that when the pipette radius or aspiration length is comparable to the elastocapillary length, surface energy may distinctly alter the aspiration response. Generally, both the aspiration length and the bulk normal stress decreases as surface energy increases, and thus neglecting the surface energy will lead to an overestimation of elastic modulus. Through dimensional analysis and numerical simulations, we provide an explicit relation between the imposed pressure and the aspiration length. This method can be applied to accurately determine the mechanical properties of soft biological tissues and organs, e.g., tumors and embryos.

**Key words**: micropipette aspiration, biological tissue, surface energy, hyperelasticity

## 1. Introduction

The mechanical characterization of cells and biological tissues is a key issue in medical engineering, e.g., disease diagnosis (1), tissue engineering, and biomimetics. In recent years, various techniques have been developed to measure the mechanical properties of cells and soft biological tissues, for examples, nanoindentation (2), magnetic beads (3), optical stretcher (4), and micropipette aspiration (5).

Due to its simplicity and accessibility, the micropipette aspiration technique has been extensively utilized to determine the mechanical properties of cells and biological materials, such as the anisotropic elasticity of blood vessel walls (6) and the viscoelasticity of cell nucleus (7). For the pipette aspiration of an incompressible half space, Theret *et al*. derived an analytical solution based on the infinitesimal strain approximation (8). Through finite element simulations of the pipette aspiration procedure, Aoki *et al*. analyzed the influences of the configuration and size of the pipette and the specimen (9). Boudou *et al*. presented a modified expression to describe the aspiration of a linear elastic material with adhesion (10). Zhao *et al*. extracted the nonlinear elastic properties of the top layer of a multilayered material from the pipette aspiration curve (11). Recently, Cao *et al*. proposed a scaling relation between the creep function and the aspiration length for viscoelastic materials and demonstrated that this relation is independent of material, geometric, and boundary nonlinearities (12). Through dimensional analysis and finite element simulations, Zhang *et al*. determined the explicit expressions between experimental responses and material properties for several typical hyperelastic materials (13).

For such soft materials as elastomers and polymeric gels, surface tension may significantly affect their morphology (14), contact mechanism (15-17), and stability (18). Also, surface tension plays a significant role in cell sorting (19), cell aggregation (20) and tissue spreading. On the surfaces of solid tumors, for example, there usually exists a thin membrane with significant tensile stress, which plays a crucial role not only in the mechanical properties but also in their biophysical behaviors of tumors. Recent examinations of micro-/nano-indentations have revealed that the elastic modulus of soft solids and cells would be highly overestimated if surface effect is neglected (21). For the aspiration of biological viscoelastic drops, Guevorkian *et al*. proposed a model to deduce surface tension, viscosity and elastic modulus (22). However, it remains unclear how surface energy affects the measurement results of a micropipette aspiration test.

This work is aimed to examine the influence of surface energy on pipette aspiration. Through a combination of dimensional analysis and finite element simulations, an explicit relation between the applied pressure and the aspiration length is derived. The work provides an efficient technique to extract the mechanical properties of cells, soft biological tissues and organs (e.g., livers, tumors and embryos).

## 2. Model

Fig. 1 illustrates the micropipette aspiration test of a hyperelastic material, e.g. soft solids and biological tissues. A given pressure $p$ is applied on the surface of the measured material through an aspiration tube with internal radius $a$ and wall thickness $b$. Let $l$ denote the induced aspiration length. Refer to a cylindrical coordinate system, $(r, \theta, z)$ as shown in Fig. 1, where the origin $O$ is located at the center on the surface of undeformed material inside the aspiration tube, and the $z$-axis is normal to the surface.

Since the aspiration problem of a compliant solid involves material, geometric, and boundary nonlinearities, it is difficult to derive an analytical solution and thus we adopt the finite element method (FEM) in this work. We carry out the numerical simulations using the commercial software ABAQUS. The incompressible neo-Hookean model is here used to describe the hyperelasticity of the soft material (23, 24). Then the strain energy density can be written as (25)

$$U = \frac{E}{6}(I_1 - 3), \tag{1}$$

where $E$ is the elastic modulus and $I_1 = \lambda_1^2 + \lambda_2^2 + \lambda_3^2$ is the first invariant of the principal stretches $\lambda_i$. The influence of surface energy is introduced by surface elements via the user subroutine UEL. The Newton-Raphson method is adopted to seek the equilibrium state of this nonlinear problem. For the FEM implementation method of surface energy, the reader can refer to Ref. (21).

The axisymmetric finite element model used in our simulations is shown in Fig. 2. The 21437 four-node bilinear axisymmetric quadrilateral hybrid reduced integration elements (CAX4RH) are used to discretize the hyperelastic solid. User-defined surface elements are attached on the top surface of the substrate and a constant surface energy density $\gamma$ is assumed on the material surface, which corresponds to a constant surface tension.

As for the boundary conditions, the symmetric boundary condition is applied on the symmetrical axis ($r=0$) and the displacement along the $z$-direction is prescribed to be zero on the bottom surface. The micropipette is modeled to be rigid and its position is fixed during the aspiration process. The friction between pipette and substrate is considered and the coefficient of friction is set as 0.5 (13). The fillets are adopted around the edges of pipette to overcome stress concentration and the radius of fillets is taken as $0.01a$. Our numerical results illustrate that, when the ratio of thickness $b$ to internal radius $a$ is larger than 0.3, the overall aspiration response is independent of the ratio $b/a$, and we consider just this case. The aspiration length $l$ varies from 0 to $a$ in the simulations of aspiration process. Convergence tests have been carried out to ensure the accuracy of computational results.

## 3. Results and discussions

To investigate the influence of surface energy on the elastic field in the soft substrate, we consider three soft silicone materials with elastic moduli $E$ = 85 kPa, 250 kPa and 500 kPa, and the corresponding surface energy densities $\gamma$ =0.032 N/m, 0.039 N/m, and 0.035 N/m, respectively (15). We use the following elastocapillary length to quantify the surface energy effect:

$$s = \frac{2\gamma}{E^*}, \qquad (2)$$

where $E^*=E/(1-v^2)$ is the plane-strain elastic modulus and $v$ is the Poisson's ratio of the material. For the aforementioned soft silicone materials, the elastocapillary lengths are 0.564 μm, 0.234 μm, and 0.105 μm, respectively. For the micropipette aspiration problem, the ratio of the elastocapillary length $s$ to the internal radius of tube $a$ is adopted to represent the relative significance of surface energy to bulk deformation.

For a specified internal pressure $p=0.5E$ and the tube thickness $b=0.5a$, the dimensionless normal displacement $w$ on the surfaces of various substrates are plotted in Fig. 3. For the cases without surface energy, the normalized surface profiles coincide with each other and thus are independent of the elastic modulus. When surface energy is included, the aspiration length becomes lower than that without surface energy, indicating a stiffening effect of surface energy. As the ratio $s/a$ rises, i.e., increasing surface effects, the normal displacement on the surface both inside and outside the pipette decreases.

Fig. 4 displays the normal Cauchy stress $\sigma_z$ on the surfaces of various substrates inside the pipette. For the cases without surface energy, the normal stress in soft solids remains almost constant as the internal pressure $p$ in most area ($r/a<0.8$), and rises rapidly when approaching the pipette edge. This characteristic is independent of elastic modulus. When surface energy is taken into account, the normal stress in the substrate becomes smaller than the internal pressure $p$ due to the capillary effect. The larger the value $s/a$ is, the stronger the capillary effect, and the more significant the decrease in the normal stress, especially around the pipette edge where the surface curvature is large.

For hyperelastic soft materials without surface energy, Zhang *et al.* has proposed explicit expressions for the pipette aspiration. For small deformation ($l/a \leq 0.3$), the applied pressure $p$ is linearly proportional to the aspiration length $l$ as

$$p_0 = 1.07 E \left( \frac{l}{a} \right), \tag{3}$$

and for large deformation ($l/a > 0.3$), the relation becomes

$$p_0 = E \left[ 0.872 \left( \frac{l}{a} \right) + 0.748 \left( \frac{l}{a} \right)^2 \right]. \tag{4}$$

When surface energy is taken into account, the above relations needed to be modified.

For different pipette internal radii and material properties, we perform finite element simulations to achieve an explicit relation between the imposed pressure and the aspiration length. In the case of small deformation ($l/a \leq 0.3$), according to dimensional analysis, the pressure $p_s$ depends not only on $l/a$ but also on $a/s$, and thus

$$p_s = E f_s \left( \frac{l}{a}, \frac{a}{s} \right). \tag{5}$$

For various specified values of $l/a$ (i.e. 0.05, 0.1, 0.15, 0.2 and 0.3), we calculate the variation in the applied pressure $p_s$ with respect to $a/s$. To determine the dependence of the applied pressure $p_s$ on $a/s$, we normalized $p_s$ by $p_0$ predicted by Eq. (3), as displayed in Fig. 5. It is seen that the normalized pressure $p_s / p_0$ depends only on the ratio $s/a$ and is independent of the ratio $l/a$. In this way, we can explicitly achieve a relation between the internal pressure and the aspiration length as

$$p_s = 1.07 E \frac{l}{a}\left(1 + 2.4 \frac{s}{a}\right) \quad (l/a \leq 0.3). \tag{6}$$

It is found that, when the internal radius $a$ of pipette is comparable to or smaller than the elastocapillary length $s$, surface energy will remarkably alter the relation between the pressure and the aspiration length. For soft materials and biological cells, the elastocapillary lengths can be up to several micrometers. Thus when the pipette radius is on the order of micrometers, the effect of surface energy on the mechanical response of aspiration will be significant. For a specialized pressure, Eq. (6) predicts a smaller aspiration length, which leads the measured material to be seemingly stiffer. When the pipette radius is much larger than the elastocapillary length, the influence of surface energy is negligible and Eq. (6) reduces to the case without surface energy, i.e., Eq. (3).

In the case of large deformation ($l/a>0.3$), our numerical results indicate that the pressure $p_s$ depends on the length ratios $l/a$ and $l/s$, and thus

$$p_s = E f_s\left(\frac{l}{a}, \frac{l}{s}\right). \tag{7}$$

To achieve the relation between the applied pressure and the aspiration length, we perform numerical simulations for the cases of $l/a=$ 0.4, 0.5, 0.7, 0.9 and 1. Based on dimensional analysis, the variation of pressure $p_s$ normalized by $p_0$ from Eq. (4) with respect to $l/s$ is displayed in Fig. 6. It can be found that, at large deformation, the normalized pressure $p_s/p_0$ depends on the ratio $s/l$. The dependence of the pressure on the aspiration length can be well approximated as

$$p_s = E\left[0.872\left(\frac{l}{a}\right) + 0.748\left(\frac{l}{a}\right)^2\right]\left[1 + 0.715\frac{s}{l}\right] \quad (l/a > 0.3). \tag{8}$$

Fig. 6 shows that the relation between the pressure and the aspiration length can be evidently altered by surface energy when the aspiration length $l$ is smaller or close to the elastocapillary length $s$. For a given aspiration length, Eq. (8) predicts a higher pressure for materials with surface energy than that without surface energy. When the aspiration length $l$ is much larger than the elastocapillary length $s$, the influence of surface energy vanishes and the pressure can be directly predicted by Eq. (4). For the case with large elastocapillary length, the explicit expressions Eqs. (6) and (8) considering surface energy can provide a more accurate avenue to evaluate the material properties from the pipette aspiration test.

## 4. Conclusions

In summary, we have investigated the micropipette aspiration method for measuring the hyperelasticity of soft materials with surface effect. It is found that surface energy can decrease the normal surface displacement and the internal pressure. Through dimensional analysis and numerical simulations, an explicit expression between the applied pressure and the aspiration length has been given. The results show that when the pipette radius or the aspiration length is comparable to the elastocapillary length, surface energy can remarkably influence the pressure-length relation. This work proposes a more accurate approach to extract the mechanical properties of soft solids and biological materials from pipette aspiration experiments.

**Acknowledgements**

Supports from the National Natural Science Foundation of China (Grant No. 11525209 and 11432008) are acknowledged.

**Figure captions:**

Fig. 1. Pipette aspiration method of hyperelastic half space.

Fig. 2. Finite element models of the pipette aspiration test.

Fig. 3. Normal displacement distribution on the surface of the substrate.

Fig. 4. Normal stress distribution inside the pipette.

Fig. 5. Variation of the load with respect to the pipette internal radius for $l/a \leq 0.3$.

Fig. 6. Dependence of the load on the aspiration length for $l/a > 0.3$.

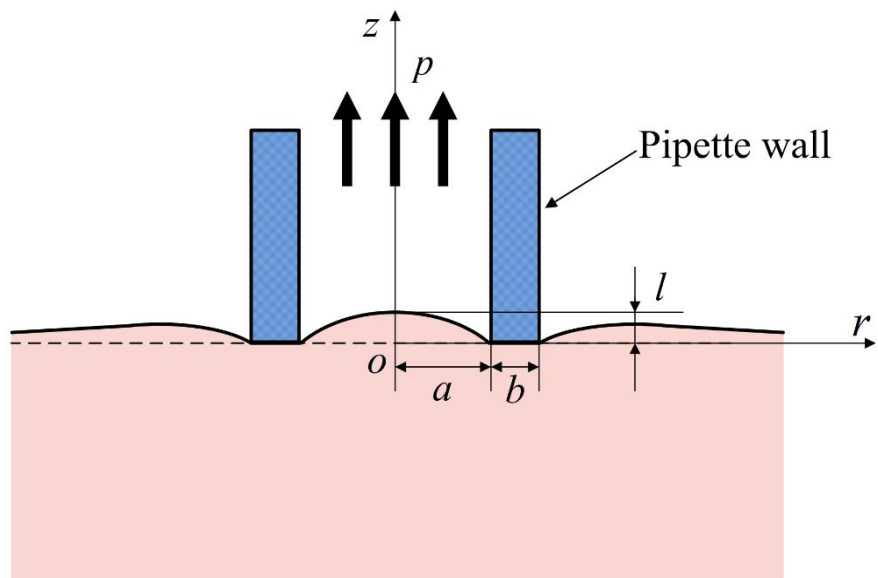

Fig. 1

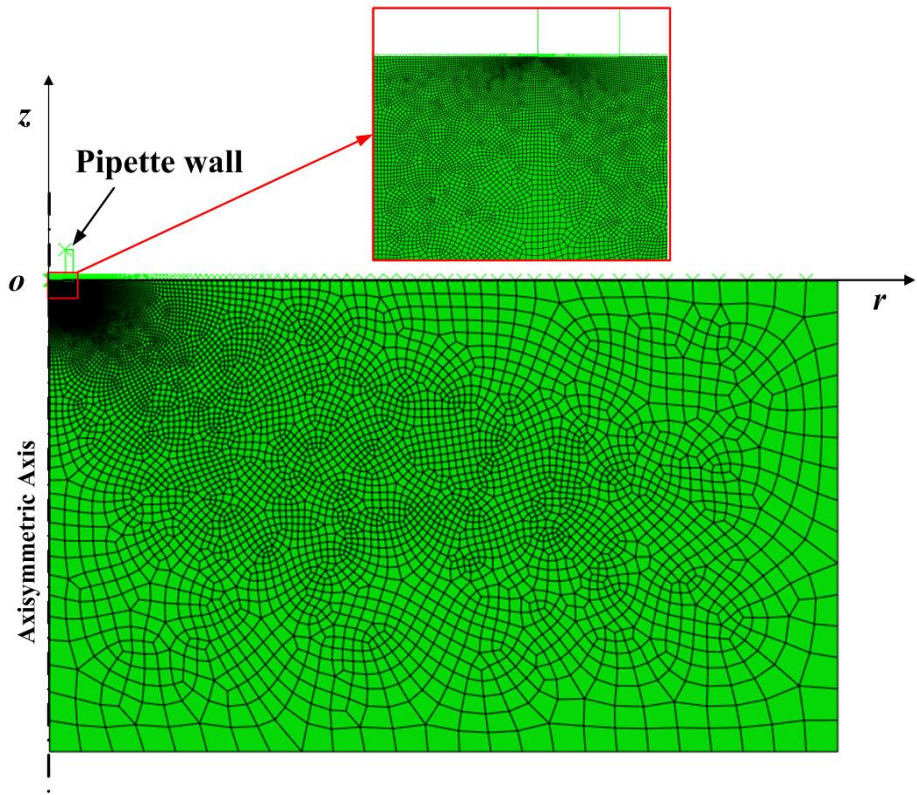

Fig. 2

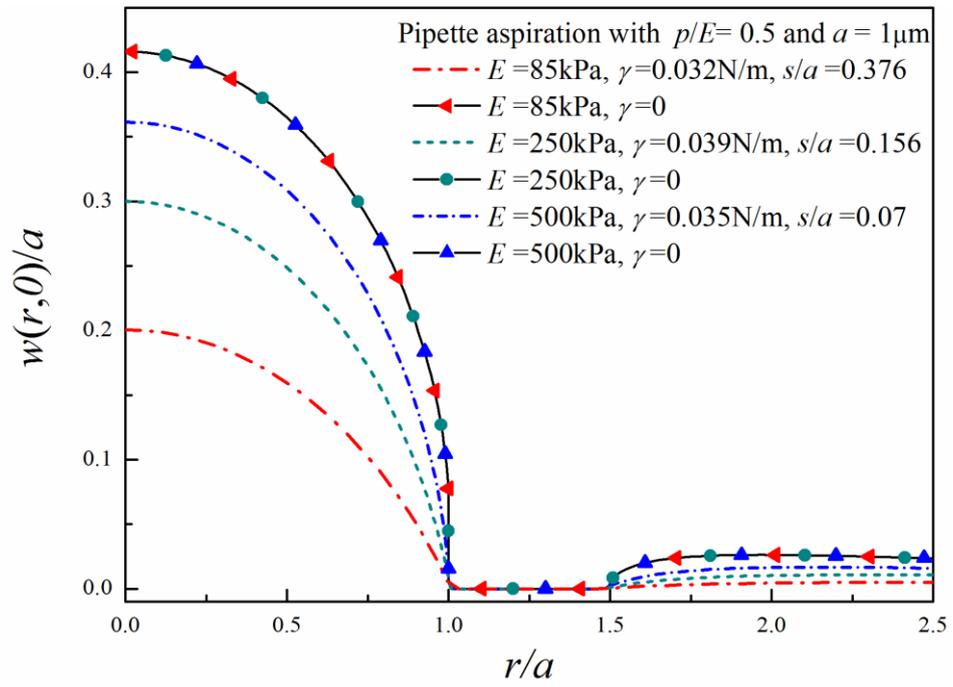

Fig. 3

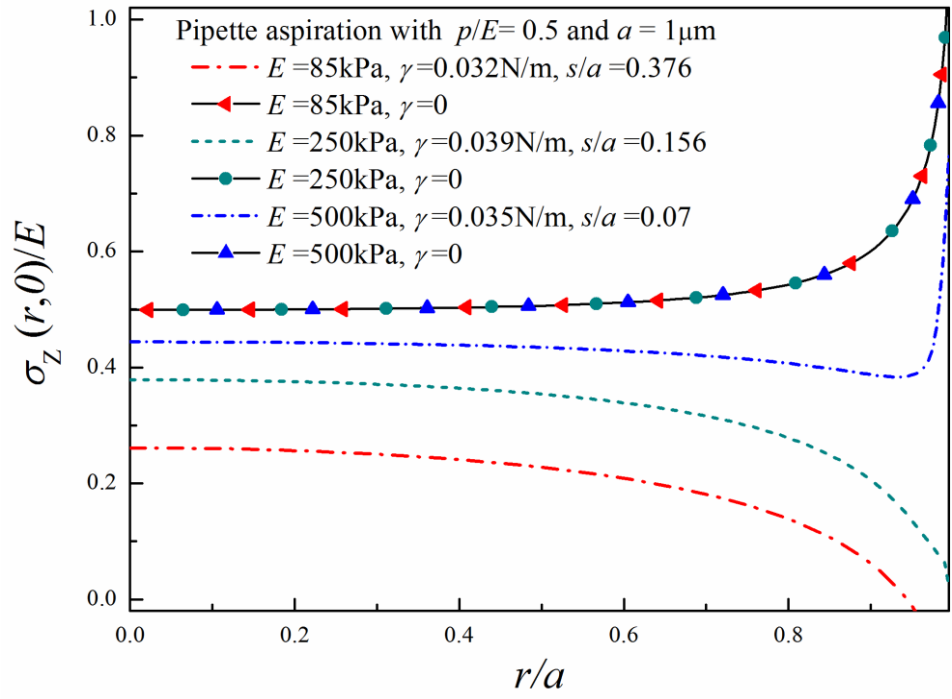

Fig. 4

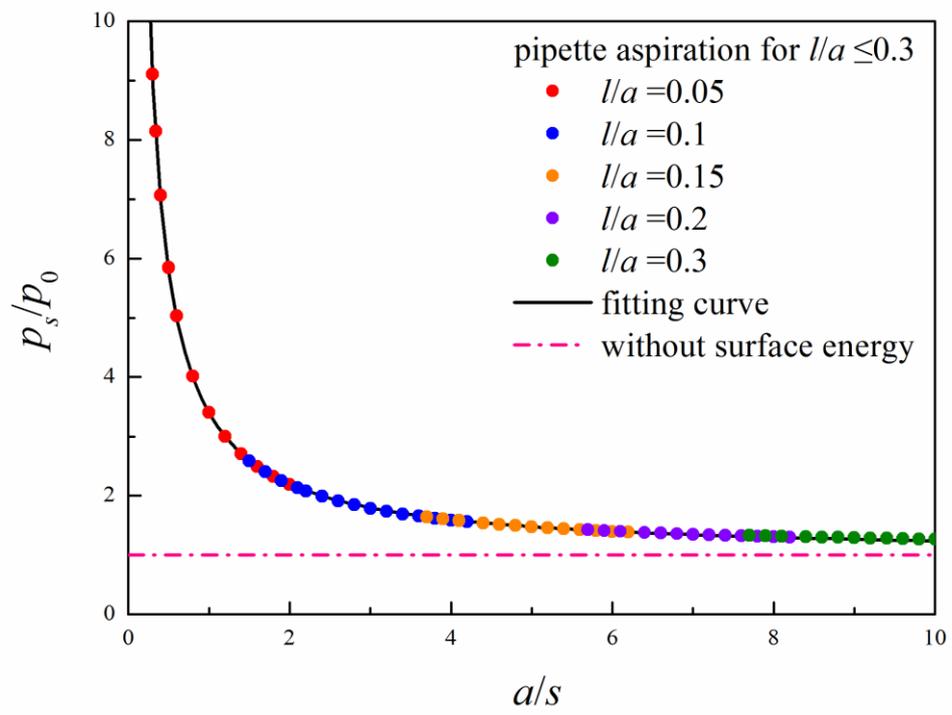

Fig. 5

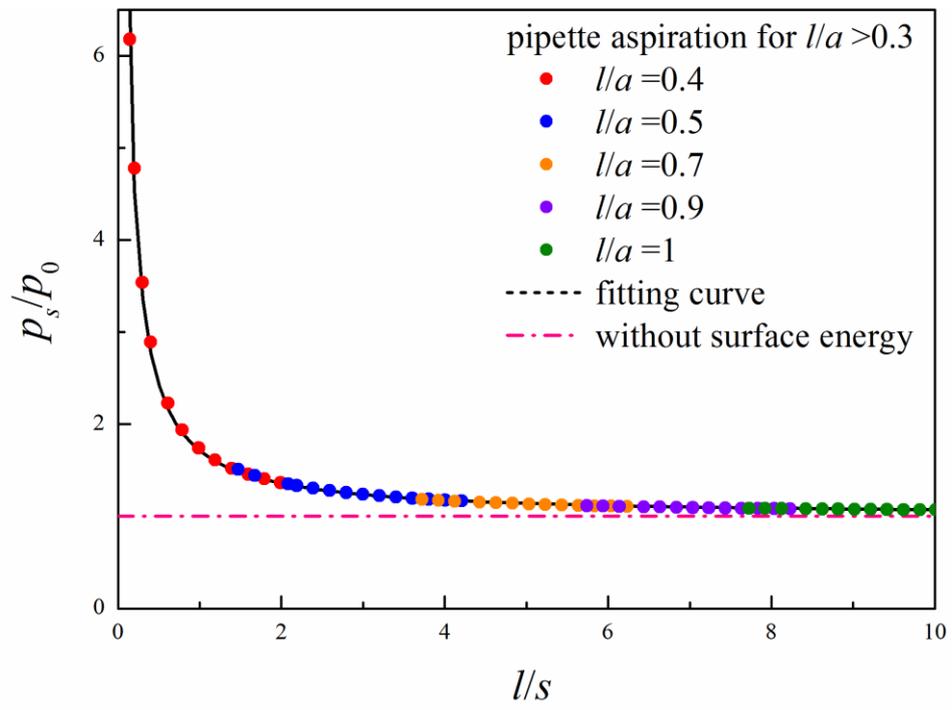

Fig. 6